\documentstyle[psfig,prb,aps]{revtex}
\draft

\title{Solution of the Schr\"odinger Equation for Quantum Dot Lattices
with Coulomb Interaction between the Dots}
\author{by M. Taut\\Institut f\"ur Festk\"orper-- und Werkstoff-- Forschung
Dresden\\ Postfach 270016,
01171 Dresden, Germany\\
email: m.taut@ifw-dresden.de\\}

\begin{document}
\maketitle

\begin{abstract}
The Schr\"odinger equation for quantum dot lattices with non-cubic,
non-Bravais lattices built up from elliptical dots is investigated.
The Coulomb interaction between the dots is considered in dipole approximation.
Then only the center of mass (c.m.) coordinates of different dots couple
with each other.
This c.m. subsystem can be solved exactly
and provides magneto-- phonon like {\em collective excitations}.
The inter--dot interaction is involved only through
 a single interaction parameter.
The relative coordinates of individual dots form decoupled subsystems
giving rise to {\em intra--dot excitation} modes.
As an example, the latter are calculated exactly for two--electron dots.\\
Emphasis is layed on {\em qualitative} effects like:
i) Influence of the magnetic field on the lattice instability
due to inter--dot interaction, ii) Closing of the gap  between
the lower and the upper c.m. mode at B=0 for elliptical
dots due to dot interaction, and iii) Kinks in the intra dot excitation
energies
(versus magnetic field) due to change of ground state angular momentum.
It is shown that
for obtaining striking qualitative effects one should
go beyond simple cubic lattices with circular dots.
In particular, 
for observing effects
of electron-- electron interaction between the dots in FIR 
spectra (breaking Kohn's Theorem) one has to consider dot lattices
with at least two dot species with different confinement tensors.
\end{abstract}

\section{Introduction}
Quantum dots have been in the focus of intensive research already for at least
a decade  which lead to a countless number of 
publications \footnote{Therefore we will refer here only to papers which
are  directly connected to the scope of this work} 
(for a recent book see Ref.\onlinecite{Hawrylak-Buch}). 
Although almost all experiments are performed at dot lattices,
in the vast majority of theoretical investigations the interaction between dots
is neglected. This is for the following reasons: 
i) Because the confinement frequency $\omega_0$ is a parameter, which
is mainly extracted from optical properties, it is difficult 
to tell the influence of dot interaction
apart from the intrinsic single-- dot value. (Possibilities to
overcome this problem are discussed in the present work.)
ii)  The theory   of Raman spectra, 
which can in principle monitor 
the dispersion (wave number dependence) of excitation energies as a
direct consequence of interdot-- interaction,
is not yet advanced enough to extract the dispersion.
iii) The lattice constant of dot arrays produced with current technologies
is so large ($>2000${\AA}) that large 
electron numbers $N$ per dot are necessary
to obtain a seizable  amount of shift.  For these N, however,
 reliable first principle calculations are not possible.
With the advent of self-- assembled dot arrays the last item might change.\\
The scope of this paper is to investigate conditions, which lead to
{\em qualitative} and observable effects of interdot-- interaction
on excitation spectra and the phase transition found in Ref.\onlinecite{Broido}.
Unlike in Ref.\onlinecite{Broido},  a magnetic field {\bf B}
 is explicitly taken into account and a microscopic theory is applied. 
Our approach is purely microscopic, i.e. we solve the Schr\"odinger
equation of a model system {\em exactly}. Our model comprises the following
approximations: i) The dot confinement is strictly parabolic in 
radial direction, but with anisotropic 
confinement frequencies $\omega_i\; (i=1,2)$ and independent of $N$ and {\bf B}.
ii) Overlap of wave functions between different dots is neglected (no hopping).
iii) The Coulomb interaction of the electrons in different dots is 
treated in dipole approximation (second order in dot diameter over lattice constant).
Our model is similar to that in Ref.\onlinecite{Halperin}, but allows more complicated
dots and lattice structures. Besides we calculate 
also the intra dot excitations 
(apart from the collective center-- of-- mass excitations)
for $N=2$ explicitly and discuss the instability \cite{Broido} in this
microscopic model. Our results on the lateral dot dimer are compared
with a former paper \cite{Chakraborty}, 
which uses a high magnetic field approach, in Sect. III. \\
The plan of this paper is as follows. For further reference, 
we briefly summarize in Sect. I 
some relevant results for one single dot, or for dot lattices, where the
distance between the dots is very large. This is important,
because all exact solutions in the center-- of-- mass subsystem 
are traced back (by special transformations) to the solution of this
one-- electron Hamiltonian. This is analogous to ordinary
molecular and lattice dynamics.
 After this, we consider a dot dimer,
which mimics a lattice, where the dots are pairwise close to each other.
This model can give an idea of the effects expected in
dot lattices with a basis. Next we consider a rectangular, but primitive
lattice in order to obtain the dispersion in the spectra. 
Finally, the intra-dot excitations of the Hamiltonian in the
relative coordinates are calculated numerically. The paper ends with a summary.
In the Appendix we give a short and elementary proof for the fact
 that the Generalized Kohn Theorem holds even for arbitrary
arrays of  identical non-circular quantum dots 
with Coulomb interaction (between the dots) in an homogeneous magnetic field.

\section{Single Dot}
The Hamiltonian considered here reads (in atomic units $\hbar=m=e=1$)
\begin{equation}
H=\sum^N_{i=1}
\left\{  
\frac{1}{2 m^*}
\left[ 
{\bf p}_i+
{1\over c} {\bf A}({\bf r}_i) 
\right]^2 +
\frac{1}{2} \;{\bf r}_i \cdot 
{\bf C }
\cdot {\bf r}_i 
\right\} 
+ \frac{1}{2} \sum_{i\ne k}\;\frac{\beta}{|{\bf r}_i-{\bf r}_k|}
\end{equation}
where $m^*$ is the effective mass (in units of the bare
electron mass $m$), $\beta$ the inverse dielectric 
constant of the background, and
$\bf C$ a symmetric tensor. In case 
of a single dot, $\bf C$ is given  by the confinement potential
and we define ${\bf C}={\bf \Omega}$.
It is always possible to find a coordinate system where
$\Omega_{12}=\Omega_{21}=0$ and $\Omega_{ii}=\omega_i^2=m^*\omega_i^{*2}$.
We use the symmetric gauge ${\bf A}=\frac{1}{2}\;{\bf B}\times {\bf r}$
throughout.
The Zeeman term
in $H$ is disregarded at the moment.\\
For $N=1$, the Hamiltonian
\begin{equation}
H=
\frac{1}{2 m^*}
\left[
{\bf p}+
{1\over c} {\bf A}({\bf r})
\right]^2 +
\frac{1}{2}\; {\bf r} \cdot
{\bf C }
\cdot {\bf r}
\label{H-N=1}
\end{equation}
can be diagonalized exactly.
Later on we will see that also the case of interacting dots
can be traced back to the solution of type (\ref{H-N=1}).
(Therefore, we kept the off diagonal elements of $\bf C$ in the results given below
because the dynamical matrix, which also contributes to $\bf C$, is
generally non--diagonal and we want to use the same coordinate system for all
$\bf q$ values.)
After transforming the
operators ${\bf r}_i$ and ${\bf p}_i$ to creation-- annihilation operators
(see e.g. Ref.\onlinecite{Hawrylak-Buch}) and using the procedure described by
Tsallis \cite{Tsallis},
we obtain for the eigenvalues
\begin{equation}
E(n_+,n_-)= (n_+ +\frac{1}{2})\; \omega_+ + (n_- +\frac{1}{2})\; \omega_- ~;
~~~n_\pm=0,1,2,...
\label{E-N=1}
\end{equation}
where
\begin{eqnarray}
\omega_\pm &=& \sqrt{
\frac{\omega_c^{*2}}{2}+\tilde\omega_0^2 \pm 
\sqrt{\frac{\omega_c^{*4}}{4}+\omega_c^{*2}\;\tilde\omega_0^2+\frac{\Delta^2}{4}
+C_{12}^2}}
\label{omega-pm-1}\\
&=&\sqrt{
\left[
\frac{1}{2} \sqrt{\omega_c^{*2}+4\; \tilde\omega_0^2+
\frac{(\Delta^2+4\;C_{12}^2)}{\omega_c^{*2}} }
\pm \frac{\omega_c^*}{2}
\right]^2
-\frac{(\Delta^2+4\;C_{12}^2)}{4\;\omega_c^{*2}}  }
\label{omega-pm-2}
\end{eqnarray}
\begin{equation}
\tilde\omega_0^2=\frac{1}{2}(C_{11}+C_{22})~;~~~~~
\Delta=C_{11}-C_{22}
\end{equation}
and $\omega_c^*=\frac{B}{m^* c}$ is the cyclotron frequency 
with the effective mass.
(The results for the special case $C_{12}=0$ can also be found in Ref. 
\onlinecite{other}.)
The optical selection rules are the same as in the circular case, i.e.,
there are two possible types of excitations 
\begin{equation}
(\Delta n_+=\pm 1 ~~~\mbox{and}~~~\Delta n_-=0)~~~\mbox{or}~~~
(\Delta n_-=\pm 1 ~~~\mbox{and}~~~\Delta n_+=0)
\end{equation}
leading to the excitation energies $\Delta E=\omega_+$ and $\omega_-$. 
It is easily seen that the form (\ref{omega-pm-2}) reduces
to the familiar formula in the circular case, where $\Delta=0$ and
$C_{12}=0$. By inspection of (\ref{omega-pm-1}) we find
that a {\em soft mode} $\omega_-(B)=0$ can only occur if
$C_{11}\cdot C_{22}=C_{12}^2$. For a diagonal $\bf C$ this means that
$min(C_{11},C_{22})=0$. The last condition is of importance for
interacting dots considered in the next Sections.\\
In  the limiting case $B = 0$ we obtain from (\ref{omega-pm-1})
\begin{equation}
\omega_\pm(B=0)=\sqrt{\frac{(C_{11}+C_{22})}{2}
\pm \sqrt{\frac{(C_{11}-C_{22})^2}{4}+C_{12}^2}}
\label{B=0}
\end{equation}
We see that {\em  degeneracy} $\omega_+(B=0)=\omega_-(B=0)$ can only happen if
$C_{12}=0$ {\em and} $C_{11}=C_{22}$.
For a diagonal confinement tensor with $C_{12}=0$ we obtain
$\omega_+(B=0)=\mbox{max}(\omega_1,\omega_2)$ and 
$\omega_-(B=0)=\mbox{min}(\omega_1,\omega_2)$. 
As to be expected, we observe a gap
between the two excitation curves $\omega_+(B)$ and $\omega_-(B)$ at
$B=0$, if the two confinement frequencies do not agree.\\
Alternatively we can introduce the quantum numbers
\begin{equation}
k=\frac{(n_+ + n_-)-|n_+ + n_-|}{2}~;~~~m_z=n_+ - n_-
\end{equation}
where $k$ is the node number and $m_z$ turns in the circular limit 
into the angular momentum quantum number.\\

For arbitrary $N$, the center of mass (c.m.) 
${\bf R}=\frac{1}{N}\sum_i {\bf r}_i$
can be separated $H=H_{c.m.}+H_{rel.}$ with
\begin{equation}
H_{c.m.}=\frac{1}{N}
\left\{
\frac{1}{2 m^*}
\left[
{\bf P}+
{N\over c} {\bf A}({\bf R})
\right]^2 +
\frac{N^2}{2} {\bf R} \cdot
{\bf C }
\cdot {\bf R}
\right\}
\label{H-cm}
\end{equation}
where ${\bf P}=-i \nabla _{\bf R}$ (see Appendix). As well known, $H_{c.m.}$
does not contain the electron-- electron-- interaction. $H_{c.m.}$ 
can be obtained from the one-- electron Hamiltonian (\ref{H-N=1}) by 
the substitution:
$B \rightarrow N B$,~${\bf C} \rightarrow N^2 {\bf C}$ and
$H \rightarrow \frac{1}{N} H$. If we make the same 
substitution in the eigenvalues (\ref{E-N=1}), we obtain 
$$E_{c.m.}(n_+,n_-)=E_{N=1}(n_+,n_-)$$
i.e., the eigenvalues of the c.m. Hamiltonian are independent of $N$.
In other words, in $H$ there are excitations, in which the pair correlation 
function is not changed, or classically speaking, where the charge distribution
oscillates rigidly.
 Because FIR radiation (in the limit
$\lambda \rightarrow \infty$) can excite only the c.m. subspace, all
we see in FIR spectra is the c.m. modes.

\section{Dot Dimer}
We consider two {\em identical} elliptical dots centered at 
$\mbox{\boldmath $a$}_1=(-a/2,0)$
and $\mbox{\boldmath $a$}_2=(+a/2,0)$.
We expand the Coulomb interaction between electrons
in {\em different} dots in a multi-pole series and restrict ourselves
to the dipole approximation.
By introduction
of c.m. and relative coordinates within each dot, the c.m. coordinates and the
relative coordinates decouple 
\footnote{ The tilde
indicates that in this preliminary Hamiltonian a common gauge
center for both dots is used.}
\begin{equation}
\tilde H=\tilde H_{c.m.}({\bf R}_1,{\bf R}_2) +
 \sum_\alpha^{1,2}\;\tilde H_\alpha\left(\{ {\bf r} \}_\alpha^{(N-1)}\right)
\label{separation}
\end{equation}
$\{ {\bf r} \}_\alpha^{(N-1)}$ symbolizes $(N-1)$ relative coordinates in
the $\alpha^{th}$ dot.
This means, we have 3 decoupled Hamiltonians: the c.m. Hamiltonian and two
Hamiltonians in the relative coordinates of either dot. 
This leads to two types of excitations:\\
 i) {\em Collective excitations }
from $\tilde H_{c.m.}$ which involve 
the c.m. coordinates of both dots simultaneously.
Because of the harmonic form (in the dipole approximation), there are exactly
 two modes per dot, thus a total of four.
 Each excitation can be classically visualized as 
vibrations of rigidly moving charge distributions of both dots.\\
ii) {\em Intra dot excitations} which are doubly degenerate 
for two identical dots. Because 
$\tilde H_\alpha(\{ {\bf r} \}_\alpha^{(N-1)})$ 
is not
harmonic (it includes the exact Coulomb interaction between the electrons
within each dot, which is not harmonic), this spectrum is very complex. It
is the excitation spectrum of a single dot {\em in a modified
confinement potential} where the c.m. coordinate is
fixed. The extra term in the modified confinement potential
comes from the dipole contribution of the interdot Coulomb interaction.\\
In this Section we consider only the c.m. Hamiltonian and focus our attention
to the the effects of ellipticity in the dot confinement potential. The 
relative Hamiltonian for $N=2$ is explicitly given in the last Section
and solved for circular dots. For the elliptical confinement potential
considered in this Section, the relative Hamiltonian cannot be solved
easily, even if we restrict ourselves to $N=2$, because the elliptic
confinement potential breaks the circular symmetry of the
rest of the relative Hamiltonian.

\subsection{ Center of Mass Hamiltonian of the Dimer}
The c.m. Hamiltonian in the dipole approximation reads 
\begin{equation}
\tilde H_{c.m.}=\frac{1}{N}
\left\{ \sum_\alpha ^{1,2}
\frac{1}{2 m^*}
\left[
{\bf P_\alpha}+
{N\over c} {\bf A}({\bf U_\alpha} + \mbox{\boldmath $a$}_\alpha)
\right]^2 +
\frac{N^2}{2} \sum_{\alpha,\alpha'}{\bf U}_\alpha \cdot
{\bf C}_{\alpha,\alpha'}
\cdot {\bf U}_{\alpha'}
\right\}
\label{H-cm-dimer}
\end{equation}
 where  the small elongation ${\bf U}_\alpha$ is defined
by ${\bf R}_\alpha=\mbox{\boldmath $a$}_\alpha+ {\bf U}_\alpha$ and
${\bf P}=-i \nabla_{\bf R}=-i \nabla_{\bf U}$.
The tensor $\bf C$ is
\begin{eqnarray}
C_{\alpha,\alpha}&=&  {\bf \Omega} + \beta N \sum_{\alpha'(\ne \alpha)}
{\bf T}(\mbox{\boldmath $a$}_{\alpha, \alpha'})   \label{C1}\\
C_{\alpha,\alpha'}&=&
-\beta N \;{\bf T}(\mbox{\boldmath $a$}_{\alpha, \alpha'})~~~
\mbox{for} ~~~\alpha \ne \alpha'
\label{C2}
\end{eqnarray}
where $\mbox{\boldmath $a$}_{\alpha, \alpha'}=
\mbox{\boldmath $a$}_{\alpha}-\mbox{\boldmath $a$}_{\alpha '}$ ,
and the dipole tensor is
\begin{equation}
{\bf T}(\mbox{\boldmath $a$})=\frac{1}{a^5}\;\left [ 3 \; \mbox{\boldmath $a$}
\circ \mbox{\boldmath $a$} - a^2\;
{\bf I} \right ]
\label{dipole-tensor}
\end{equation}
containing a dyad product ($\circ$) and the unit tensor $\bf I$.
As in the c.m. system of a single dot, the explicit $N$-- dependence in
(\ref{H-cm-dimer}) cancels in the eigenvalues.
What is left is only the $N$-- dependence in the
dipole contribution of the dot interaction appearing
in (\ref{C1}) and (\ref{C2}). This means,
that the c.m. spectrum of interacting dots
is no longer independent of $N$.\\
The term $\mbox{\boldmath $a$}_\alpha$ in the argument of the
vector potential in (\ref{H-cm-dimer}) causes trouble in finding the
eigenvalues. This shift is a consequence of the fact that we
have to adopt a common gauge center for both dots 
(we chose  the middle between both dots). This problem can be
solved by applying the following unitary transformation
\begin{equation}
 H_{c.m.} = Q^{-1}\; \tilde H_{c.m.} \; Q~;~~~
Q=\prod_\alpha^{1,2}\;
e^{-i\frac{N}{2c}\;({\bf B}\times
 \mbox{\boldmath $a$}_\alpha)\cdot {\bf U}_\alpha}
\label{gauge-shift}
\end{equation}
In other words, $H_{c.m.}$ agrees with $\tilde H_{c.m.}$ except for
the missing shift in the argument of the vector potential.\\
The 4 modes inherent in $H_{c.m.}$ are not yet explicitly known, because the
4 degrees of freedom are coupled. Decoupling into two oscillator 
problems of type (\ref{H-N=1}) can be achieved by the following transformation:
\begin{equation}
{\bf U}^{(+)}=\frac{1}{2}({\bf U}_2+{\bf U}_1)~;~~~~~
{\bf U}^{(-)}={\bf U}_2-{\bf U}_1
\label{cm-diff-coord}
\end{equation}
This results in 
\begin{equation}
H_{c.m.}=\frac{1}{2} H^{(+)}\; + \; 2 H^{(-)}
\end{equation}
where
\begin{eqnarray}
H^{(+)}&=& \frac{1}{N} \left\{
\frac{1}{2 m^*}
\left[
{\bf P}^{(+)}+
{2 N\over c} {\bf A}\left({\bf U}^{(+)}\right)
\right]^2 +
\frac{N^2}{2} {\bf U}^{(+)} \cdot
\Bigg(
4   {\bf \Omega}
\Bigg)
\cdot {\bf U}^{(+)} \right\}
\label{H+}\\
H^{(-)}&=& \frac{1}{N} \left\{
\frac{1}{2 m^*}
\left[
{\bf P}^{(-)}+
{N\over 2 c} {\bf A}\left({\bf U}^{(-)}\right)
\right]^2 +
\frac{N^2}{2} {\bf U}^{(-)} \cdot
\Bigg(
\frac{1}{4} {\bf \Omega} +\frac{N}{2} \beta \; {\bf T}(\mbox{\boldmath $a$}) 
\Bigg)
\cdot {\bf U}^{(-)}) \right\}
\label{H-}
\end{eqnarray}
and $\mbox{\boldmath $a$}$ is a vector pointing from one dot center 
to the other. Then ${\bf T}(\mbox{\boldmath $a$})$ has the following components
\begin{equation}
T_{11}=\frac{2}{a^3}~;~~~T_{22}=-\frac{1}{a^3}~;~~~
T_{12}=T_{21}=0
\end{equation}
Now we assume that the principle axes of the confinement potentials
are in x-y-direction. This means
\begin{equation}
\Omega_{11}=\omega_1^2~;~~~\Omega_{22}=\omega_2^2~;~~~ 
\Omega_{12}=\Omega_{21}=0
\end{equation}
The eigenvalues of $H^{(+)}$ 
can be obtained from (\ref{E-N=1}) and (\ref{omega-pm-1})
with
\begin{equation}
\tilde\omega_0^2=\frac{1}{2}(\omega_1^2+\omega_2^2)~;~~
~\Delta=(\omega_1^2-\omega_2^2)
\label{input1}
\end{equation}
and for $H^{(-)}$ with 
\begin{equation}
\tilde\omega_0^2=\frac{1}{2}(\omega_1^2+\omega_2^2)+\frac{1}{2}\;p~;~~
~\Delta=(\omega_1^2-\omega_2^2)+3\; p
\label{input2}
\end{equation}
where the interaction parameter is defined by
\begin{equation}
p=\frac{2N\beta}{a^3}
\label{p}
\end{equation}
Observe that the dependence on $N$ cancels,
except that included in $p$ (see discussion following (\ref{H-cm})).

 It is important that the dot interaction influences
the result only through a single parameter. This conclusion
agrees with the semi-- phenomenological theory in Ref.\onlinecite{Broido}.\\

In all our figures
we express  frequencies in units of the average confinement frequency
$\omega_0=\frac{1}{2}(\omega_1+\omega_2)$, and $\Delta$ and $p$ in units
$\omega_0^2$.
Then, all
systems can be characterized by the two parameters:
 $\omega_1/\omega_2$ and  $p$.
In other words, all systems  having the $\omega_1/\omega_2$ ratio 
indicated in the figures 
are represented by the family of curves with the $p$ values shown.
The only exception we made is the cyclotron frequency $\omega^*_c$.
$\omega^*_c/\omega_0$ would be a good parameter in this sense, but we chose
to use the magnetic field in $Tesla$ instead for better physical
intuition. The conversion between both scales is given by
$\omega^*_c[a.u.^*]=\frac{0.9134\cdot10^{-2}}{m^*}\; B[Tesla]$ or,
$\omega_c^*[\omega_0]=\frac{0.9134\cdot10^{-2}}{m^*\;\omega_0[a.u.^*]}B[Tesla]$
In this paper we used $\omega_0=0.2 \;a.u.^*= 2.53\; meV$
and $m^*$ of GaAs. (We want to stress
that this choice effects only  the magnetic field scale
 and not  the qualitative features of the figures.)
For easy comparison with experimental parameters we add the definition of
effective atomic units ($a.u.^*$) 
in GaAs ($m^*=0.067, \;\beta=1/12$) for the  energy: 
$1\;a.u.^*=4.65\cdot10^{-4}\; double\; Rydberg=12.64\;meV$ ,
 and for
the length: $1\;a.u.^*=1.791\cdot10^{2}\; Bohr\; radii
=0.9477\cdot10^{2}\;\AA$.\\

Because $\bf U^{(+)}$  agrees with the
{\em total} c.m. ${\bf R}=\frac{1}{2}({\bf R}_1+{\bf R}_2)$ of the system,
$H^{(+)}$ is the total c.m. Hamiltonian. 
For $B=0$, the eigenmodes can be visualized by classical oscillations.
The  two eigenmodes of $H^{(+)}$ are  (rigid) in-- phase
 oscillations  of the dots in x and y direction, respectively. 
Because of the Kohn theorem (see Appendix), the independence of
$H^{(+)}$ on the Coulomb interaction does not only hold in the dipole 
approximation, but it is rigorous. This shows also that the dipole
approximation is consistent with the Kohn theorem, which is not guaranteed
for single particle approaches. Because FIR radiation excites 
(in the dipole approximation) only
the c.m. modes, it is only the $p$--independent 
eigenmodes of $H^{(+)}$ which are seen in 
FIR absorption experiments.\\
This statement is in contradiction
to Ref.\onlinecite{Chakraborty}. They performed numerical diagonalizations
for a lateral pair of circular dots 
confining the set of basis
functions to the lowest Landau level and considering parallel spin 
configurations only.
This is justified in the limit of high magnetic fields. They found a splitting
of the two dipole allowed modes at $B=0$ due to dot interaction and
some anti-crossing structures in the upper mode, whereas the lower mode is
always close to the single particle mode. This fact is already a strong
 indication that
the missing higher Landau levels cause both spurious effects. (Observe that the
lifting of the degeneracy at $B=0$ in the dipole allowed excitations 
in Fig.1 is due to the ellipticity
of the intrinsic confinement and not due to dot interaction.)\\
The eigenvalues of $H^{(-)}$
do depend on $p$ because the dots oscillate (rigidly)
in its two eigenmodes
in opposite phase, one mode in $x$ and one mode in $y$ direction. 
This leads to a change in the Coulomb energy. The two eigenmodes of $H^{(-)}$
can also be described as a breathing mode (in x direction) 
and a shear mode (in y direction).\\

\newpage

\begin{figure}[th]
\vspace{-2cm}
\begin{center}
{\psfig{figure=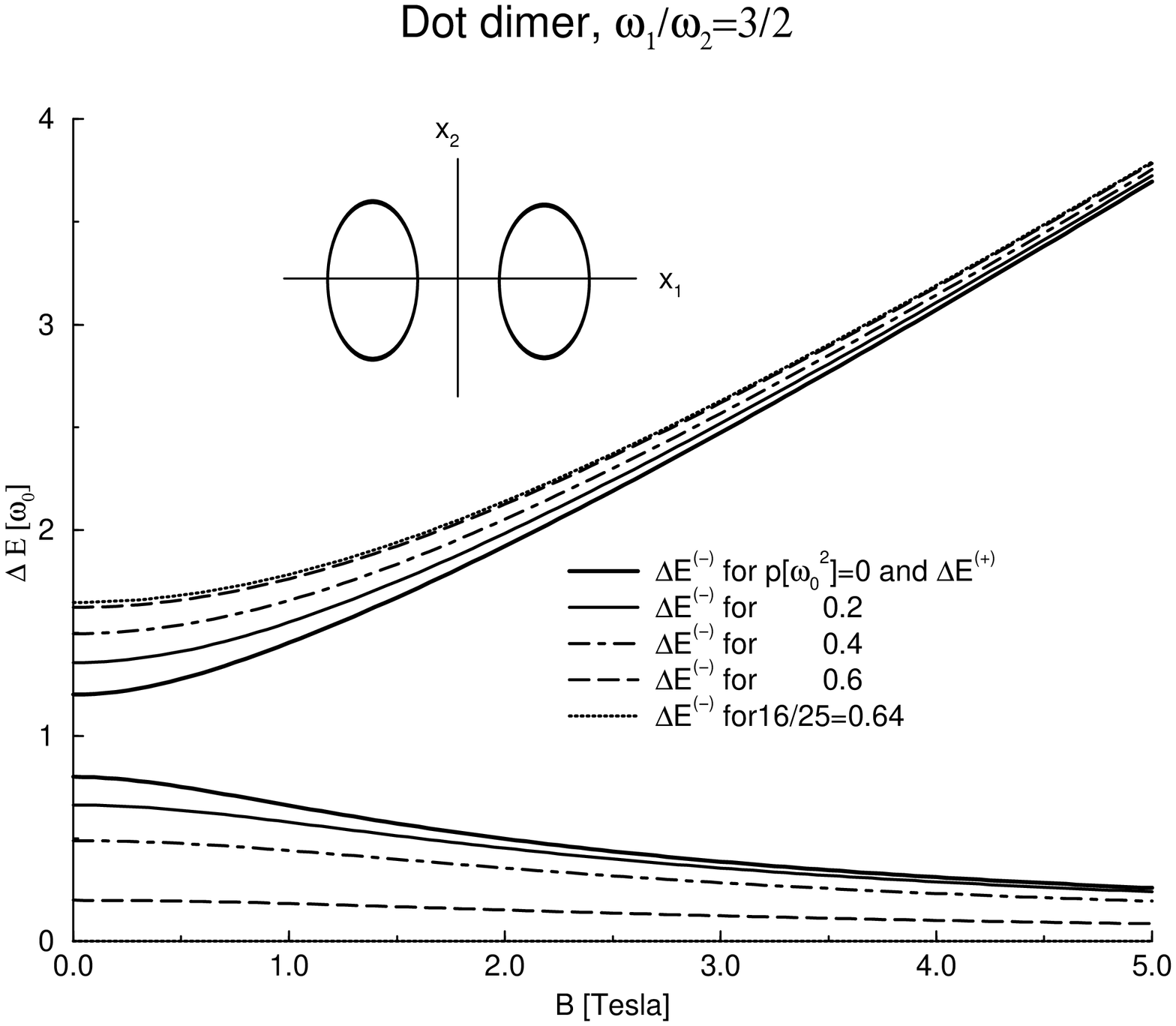,angle=-90,width=12.cm,bbllx=15pt,bblly=45pt,bburx=580pt,bbury=750pt}}
{\psfig{figure=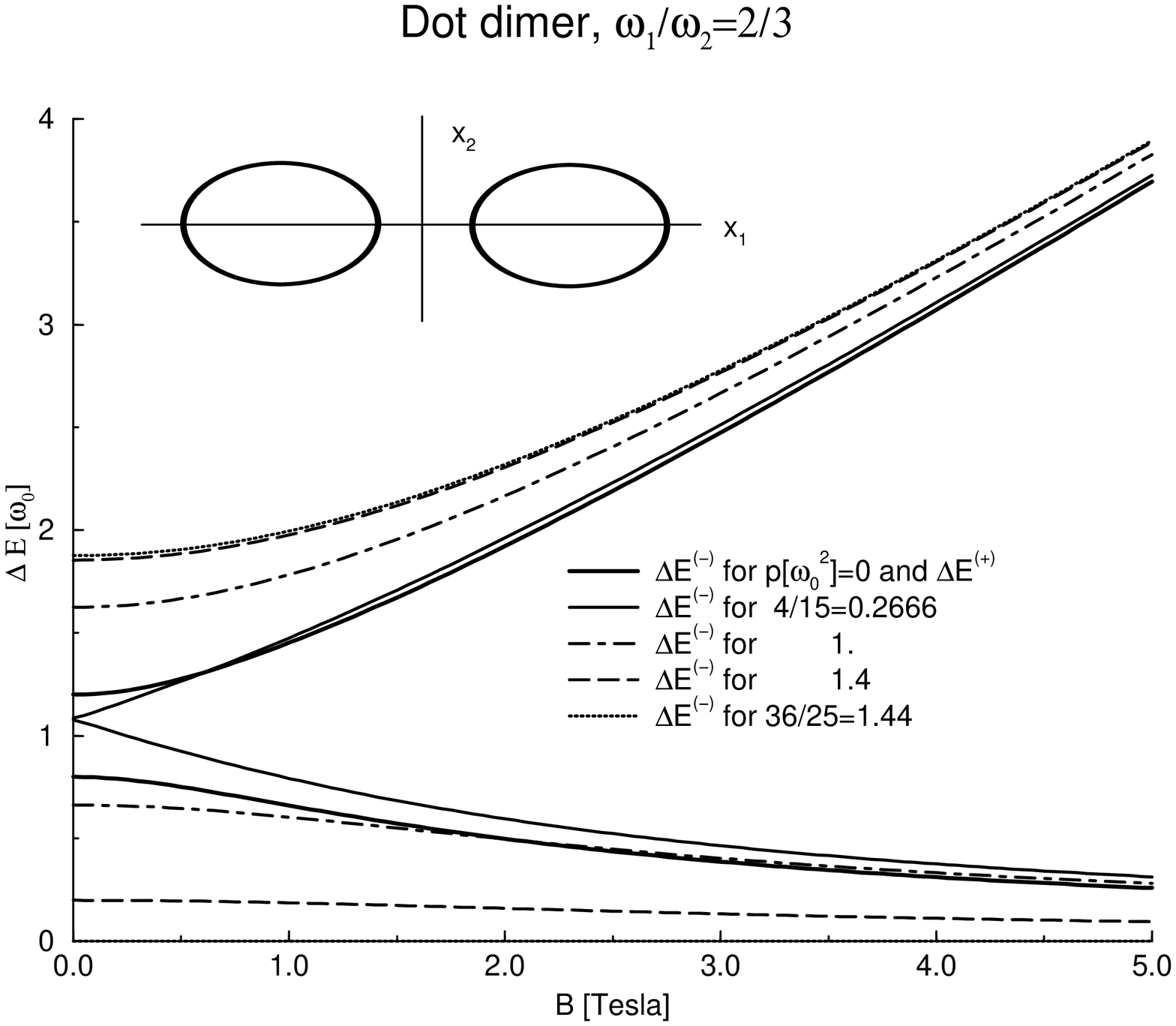,angle=-90,width=12.cm,bbllx=15pt,bblly=45pt,bburx=580pt,bbury=750pt}}
\caption[]{
Excitation energies $\Delta E^{(\pm)}_\pm =\omega^{(\pm)}_\pm$
for a dimer of elliptical dots as a
function of the magnetic field for some discrete values of the interaction
parameter $p$ ($p$ values in the inset are in units $\omega_0^2$).
The ratio of the oscillator frequencies in the direction of the
capital axes $\omega_1/\omega_2$ is a) $3/2$ and b) $2/3$.
The dipole allowed excitations $\Delta E^{(+)}$ of $H^{(+)}$ (thick full line)
are not influenced by the dot interaction and therefore independent of $p$.
}
\label{Fig1}
\end{center}
\end{figure}

\subsection{Special Features of the Excitation Spectrum}
In Fig.1a and 1b, the four excitation frequencies of the dimer are shown
with $p$ as a parameter.

For $p=0$, the two modes $\omega^{(-)}_\pm$ agree with 
the two modes $\omega^{(+)}_\pm$. 
In all symbols, the superscript sign refers to the system $H^{(+)}$
and $H^{(-)}$ (c.m. or relative
coordinate), and the subscript sign discriminates the two modes of the same 
system.
The two modes $\omega^{(+)}_\pm$ are independent of $p$.
There are two qualitatively different cases.
(Consider that $\omega_1$ is the oscillator frequency parallel to the
line, which connects the two dot centers, and $\omega_2$ is the 
oscillator frequency perpendicular to this line.)
If $\omega_1 \ge \omega_2$
 (Fig.1a), the gap between  
$\omega^{(-)}_+$ and $\omega^{(-)}_-$  at $B=0$ increases steadily
with increasing $p$   until, for a critical $p_{cr}=\omega_2^2$
(in our numerical case: $p_{cr}[\omega_0^2]=16/25=0.64$)
the lower mode $\omega^{(-)}_-$
becomes soft. This transition is {\em independent} of $B$. For 
$\omega_1 \le \omega_2$ the gap 
between  $\omega^{(-)}_+$ and $\omega^{(-)}_-$  at $B=0$
 first decreases with increasing $p$
until it vanishes for $p=\frac{1}{3}(\omega_2^2-\omega_1^2)$
 (in our numerical case: $p[\omega_0^2]=4/15=0.27$). Afterwards,
it increases  until the lattice becomes soft at $p_{cr}=\omega_2^2$
(in our numerical case: $p_{cr}[\omega_0^2]=36/25=1.44$).
The dependence of the two excitation energies
$\omega^{(-)}_+$ and $\omega^{(-)}_-$  on $p$ for $B=0$  in the second case
is shown in Fig.2. Comparison of Fig.s 2a and 2b demonstrates 
that the dot architecture
in Fig.2a is much more sensitive to interdot interaction than that in
Fig.2b. Thus, if we want to observe or use the instability,
 this event happens in case 2a for for smaller $p$
(or larger lattice constants) than in case 2b. Additionally, the assumption of
non--overlapping dot wave functions  (for a given lattice constant)
is better fulfilled in case 2a than in case 2b. \\[.5cm]

\begin{figure}[!th]
\vspace{-2cm}
\begin{center}
{\psfig{figure=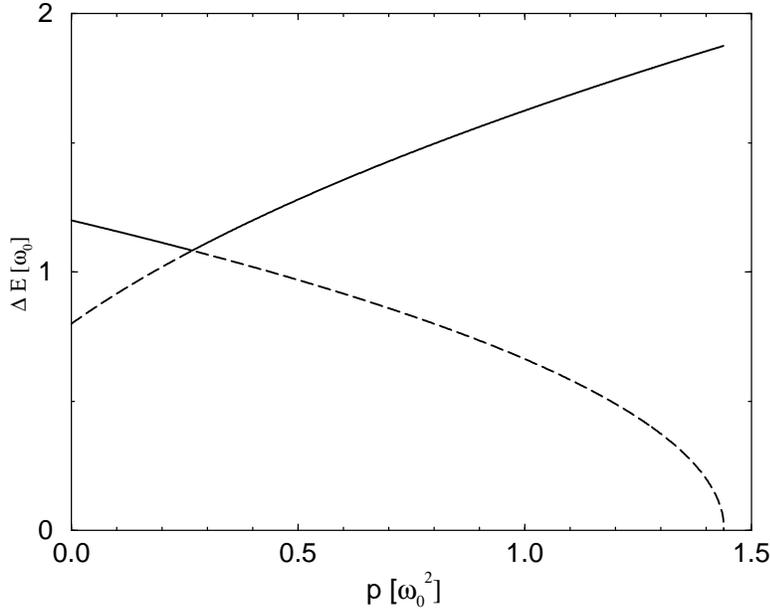,angle=-90,width=12.cm,bbllx=15pt,bblly=45pt,bburx=580pt,bbury=750pt}}
\caption[ ]{
Excitation energies $\Delta E^{(-)}_\pm =\omega^{(-)}_\pm$ of the
Hamiltonian $H^{(-)}$ for $B=0$ and $\omega_1/\omega_2=2/3$ as a function
of the interaction parameter $p$.
}
\label{Fig2}
\end{center}
\end{figure}

For GaAs as a typical substance, (\ref{p}) can be
rewritten in more convenient units as
\begin{equation}
p[\omega_0^2]=\frac{2.26 \cdot10^7 \; N}{\big(a[\AA\big])^3\;
\big(\omega_0[meV]\big)^2}
\label{p-GaAs}
\end{equation}
Obviously, we need large dots (large $N$, small $\omega_0$-- which means
large polarizability), and a small
dot distance $a$ for a seizable interaction effect.
On the other hand, the dot radius
for $N=1$  is of order of the effective magnetic length \cite{Hawrylak-Buch}
$l_0=\bigg( (2\omega_0)^2+(\omega_c^*)^2 \bigg)^{-1/4}$,
which reads  for GaAs
\begin{equation}
l_0[\AA]=\frac{238}{\bigg(\big(\omega_0[meV]\big)^2+0.739 \;
\big(B[Tesla]\big)^2\bigg) ^\frac{1}{4}}
\label{l0-GaAs}
\end{equation}
and we need small dots and high magnetic fields for small overlap.
Consequently, a magnetic field helps avoiding overlap of the dots, although
e.g. the critical $p$ for soft modes is independent of $B$.
The question is, if there exists a window between these two (partly)
conflicting demands. For an order-- of-- maigntude estimate, 
let us consider  GaAs with $\omega_0$
as chosen above and the worst case $N=1$.
Then (\ref{p-GaAs}) with a typical $p[\omega_0^2]=0.1$
(which seems to be the minimum for any observable effect)
provides  a dot distance of $a[\AA]=327$ and (\ref{l0-GaAs}) gives
for $B=0$ a radius of $l_0[\AA]=150$ and for $B[Tesla]=10$ a radius of
$l_0[\AA]=80$. Consequently, 
the constraint $l_0<a/2$ for our model can be fulfilled. 
For obtaining larger interaction effects the parameters have to be optimized.

The next question is what happens in 
{\em mode softening} physically? 
Firstly, it is the antisymmetric shear mode $\omega^{(-)}_-$
 which has the lowest frequency
and which becomes soft.
If the interaction parameter is strong enough ($p>p_{cr}$),
the {\em de}crease in interdot-- Coulomb energy
with increasing elongation of the dots
becomes larger than the {\em in}crease of confinement potential energy. Because
{\em in the harmonic model} 
both energies depend quadratically on elongation, 
the dimer {\em would} be ionized, i.e.
stripped of the electrons. Clearly, in this case we have to
go beyond the dipole approximation for the interdot interaction and 
beyond the harmonic approximation for the confinement potential.

In order to obtain a hand-waving picture of what happens,
the confinement potential of the system for shear mode oscillations 
is supplemented by a $4^{th}$ order term in the following way:
$
V_{conf.}=2 N \left[ \frac{1}{2} \omega_2^2\; U^2-A \;U^4\right]
$
with $(A > 0)$, 
and the Coulomb interaction in $4^{th}$ order reads:
$
V_{int}=-p N\; U^2+ (3pN/a^2)\; U^4
$
where $p=2 N \beta/a^3$ as above. Then, the stability condition reads
\begin{equation}
\frac{V_{tot}}{N}=(\omega_2^2-p)\; U^2+(\frac{3p}{a^2}-2A)\; U^4 \ge 0
\end{equation}
The condition for the existence of a bound state
is that the $U^4$- term is positive: $3 p/a^2>2A$.
For a positive $U^2$- term ( $p<\omega_2^2$ ), the 
equilibrium position is $U_0=0$. If the $U^2$- term becomes 
negative ( $p>\omega_2^2$ ),
the system finds a new equilibrium at a finite elongation
\begin{equation}
U_0=\pm \sqrt{\frac{(p-\omega_2^2)}{2(\frac{3p}{a^2}-2 A)}}
\end{equation}
This new ground state is doubly degenerate:
${\bf U}_1=(-a,+U_0)~,~{\bf U}_2=(+a,-U_0)$ and
${\bf U}_1=(-a,-U_0)~,~{\bf U}_2=(+a,+U_0)$ have the same energy.
In short, at $p_{cr}=\omega_2^2$ there is an electronic phase transition to
a polarised state, where the equilibrium position of the c.m. is no more in
the middle of the dots.
At the end we want to stress that all these stability considerations
are only valid if the confinement potential is not changed under
elongation of the c.m. of the dots. Secondly, it is not rigorous
to include the fourth order terms {\em after} separation of c.m. and
relative coordinates, because in fourth order these two coordinates 
do not decouple exactly.

\section{Dot Lattice}

We consider a periodic lattice of {\em equal} quantum dots at lattice sites
${\bf R}_{n,\alpha}^{(0)}={\bf R}_n^{(0)}+\mbox{\boldmath $a$}_\alpha$.
The vectors ${\bf R}_n^{(0)}$ 
form a Bravais lattice and $\mbox{\boldmath $a$}_\alpha$
runs over all sites within an unit cell.
In developing a theory for these lattices we have to repeat all steps
in Sect. II from (\ref{separation}) to (\ref{gauge-shift}) just by supplementing
the index $\alpha$ by the index $n$ for the unit cell.

\subsection{Center of Mass Hamiltonian of the Dot Lattice}
The  c.m. Hamiltonian in the dipole approximation then reads
\begin{eqnarray}
H_{c.m.}&=&\frac{1}{N}
\bigg\{ \sum_{n,\alpha} 
\frac{1}{2 m^*}
\left[
{\bf P}_{n,\alpha}+
{N\over c} {\bf A}\left({\bf U}_{n,\alpha} \right)
\right]^2  \nonumber \\
 && + \frac{N^2}{2} \sum_{n,\alpha\atop n',\alpha'}\; {\bf U}_{n,\alpha} \cdot
{\bf C}_{n,\alpha;\, n',\alpha'}
\cdot {\bf U}_{n',\alpha'}
\bigg\}
\label{H-cm-latt}
\end{eqnarray}
where ${\bf U}_{n,\alpha}={\bf R}_{n,\alpha}-{\bf R}_{n,\alpha}^{(0)}$ 
is the elongation of the c.m. 
at lattice site $(n,\alpha)$ and the force constant 
tensor $\bf C$ is defined in analogy to (\ref{C1}) and (\ref{C2}).

The Hamiltonian (\ref{H-cm-latt}) 
is a phonon Hamiltonian  in an additional homogeneous magnetic field.
The first stage of decoupling can be achieved by the usual 
phonon transformation
\begin{eqnarray}
{\bf U}_{n,\alpha}&=&\frac{1}{\sqrt{N_c}} \sum_{\bf q}^{BZ}
e^{-i{\bf q}\cdot R_n^{(0)}}\;{\bf U}_{{\bf q},\alpha}\\
{\bf P}_{n,\alpha}&=&\frac{1}{\sqrt{N_c}} \sum_{\bf q}^{BZ}
e^{+i{\bf q}\cdot R_n^{(0)}}\;{\bf P}_{{\bf q},\alpha}
\label{phonon-trafo}
\end{eqnarray}
where $N_c$ is the number of unit cells and the 
transformed coordinates have the following properties
${\bf U}_{-{\bf q},\alpha}={\bf U}_{{\bf q},\alpha}^*
={\bf U}_{{\bf q},\alpha}^\dagger$ and
${\bf P}_{-{\bf q},\alpha}={\bf P}_{{\bf q},\alpha}^\dagger$. 
The Hamiltonian in the new coordinates
is a sum of $N_c$ subsystems of dimension $2\times$ number of dots per unit 
cell: 
$H_{c.m.}= \sum_{\bf q}\;H_{\bf q}$, where
\begin{eqnarray}
H_{\bf q}&=& \frac{1}{N} \bigg\{  \sum_\alpha\; \frac{1}{2 m^*} 
\left[
{\bf P}_{{\bf q},\alpha}+\frac{N}{c}{\bf A}({\bf U}_{{\bf q},\alpha}^*)
\right]^\dagger \cdot 
\left[
{\bf P}_{{\bf q},\alpha}+\frac{N}{c}{\bf A}({\bf U}_{{\bf q},\alpha}^*)
\right] \nonumber\\
& &+ \frac{N^2}{2} \sum_{\alpha,\alpha'}
{\bf U}_{{\bf q},\alpha}^* \cdot
{\bf C}_{{\bf q};\alpha,\alpha'}
\cdot {\bf U}_{{\bf q},\alpha'}  \bigg\}
\label{H-phonon}
\end{eqnarray}
The dynamical matrix is defined by
\begin{equation}
{\bf C}_{{\bf q};\alpha,\alpha'}=\sum_n\;
e^{i{\bf q}\cdot {\bf R}_n^{(0)}}\;
{\bf C}_{\alpha,\alpha'}\left({\bf R}_n^{(0)}\right)
~;~~~~{\bf C}_{\alpha,\alpha'}\left({\bf R}_n^{(0)}\right)
= {\bf C}_{n,\alpha;\,0,\alpha'}
\label{dyn-mat-full}
\end{equation}
and it is hermitean ${\bf C}_{{\bf q};\alpha',\alpha}=
{\bf C}^*_{{\bf q};\alpha,\alpha'}={\bf C}_{-{\bf q};\alpha,\alpha'}$.\\
Next we want to recover the limiting case considered in Sect. III.
If the dots in a given unit cell are far away from those in neighboring
cells, then in (\ref{dyn-mat-full}) only the term with ${\bf R}_n^{(0)}=0$
contributes, $\bf C$ does not depend on $\bf q$, consequently the index
$\bf q$ is redundant, and (\ref{H-phonon}) agrees with (\ref{H-cm-dimer}).\\
Our preliminary result 
(\ref{H-phonon}) is not yet diagonal in $\alpha,\alpha'$. 
In some special cases (see e.g. two identical dots
per unit cell  considered in Sect. III) this can be 
achieved by an unitary transformation
\begin{equation}
{\bf U}_{{\bf q},\alpha}=\sum_{\alpha'}\;Q_{{\bf q};\alpha,\alpha'}\cdot
\tilde {\bf U}_{{\bf q},\alpha'}~;~~~~~Q^*_{{\bf q};\alpha',\alpha}=
Q^{-1}_{{\bf q};\alpha,\alpha'}
\end{equation}
under which the one-- particle term in (\ref{H-phonon}) is invariant
and the transformed interaction term 
\begin{equation}
\frac{1}{2} \sum_{\alpha,\alpha'}
\tilde{\bf U}_{{\bf q},\alpha}^* \cdot
\tilde{\bf C}_{{\bf q};\alpha,\alpha'}
\cdot \tilde{\bf U}_{{\bf q},\alpha'}~~~\mbox{with}~~~
\tilde{\bf C}_{{\bf q};\alpha,\alpha'}= \sum_{\alpha_1,\alpha_2}
Q^{-1}_{\alpha,\alpha_1}\;{\bf C}_{{\bf q};\alpha_1,\alpha_2}\;
Q_{\alpha_2,\alpha'}
\end{equation}
can be made diagonal $\tilde{\bf C}_{{\bf q};\alpha,\alpha'}=
\tilde{\bf C}_{{\bf q};\alpha}\;\delta_{\alpha,\alpha'}$ by 
a proper choice of $Q_{\alpha,\alpha'}$. Now, (\ref{H-phonon}) reads 
$H_{\bf q}=\sum_{\alpha}\;H_{{\bf q},\alpha}$, where
\begin{eqnarray}
H_{{\bf q},\alpha}&=& \frac{1}{N} \bigg\{ \frac{1}{2 m^*}
\left[
\tilde{\bf P}_{{\bf q},\alpha}+\frac{N}{c}{\bf A} (
\tilde{\bf U}_{{\bf q},\alpha}^*)
\right]^\dagger \cdot
\left[
\tilde{\bf P}_{{\bf q},\alpha}+\frac{N}{c}{\bf A} (
\tilde{\bf U}_{{\bf q},\alpha}^*)
\right] \nonumber\\
&&+ \frac{N^2}{2} 
\tilde{\bf U}_{{\bf q},\alpha}^* \cdot
\tilde{\bf C}_{{\bf q};\alpha}
\cdot \tilde{\bf U}_{{\bf q},\alpha} \bigg\}
\label{H-phonon-diag}
\end{eqnarray}
The eigenvalues of (\ref{H-phonon-diag}) can be obtained from those
of (\ref{H-N=1}) because corresponding quantities have the same
commutation rules. Such an unitary transformation  does not
exist, e.g., for two different dots per cell. Then (\ref{H-phonon})
has to be solved directly using the method described in Ref. 
\onlinecite{Tsallis}.\\

\subsection{Dynamical Matrix for Bravais lattices}
From now on we consider {\em Bravais lattices} what means that we can forget
the indices $\alpha$  in the first part of this Section. 
Then the dynamical matrix 
\begin{equation}
{\bf C}_{\bf q}={\bf \Omega}+\beta N \;\sum_{{\bf R}_n^{(0)}\ne0}
\left( 1-e^{i{\bf q}\cdot {\bf R}_n^{(0)}} \right) \;
{\bf T}\left({\bf R}_n^{(0)}\right)
\label{dyn-mat}
\end{equation}
is real and symmetric, but generally not diagonal, even if $\bf \Omega$
is diagonal.
A very important conclusion is apparent in (\ref{dyn-mat}). {\em In the limit
${\bf q}\rightarrow 0$, the inter-- dot interaction (represented by $\beta$) 
has no influence on ${\bf C}_{\bf q}$
and therefore on the spectrum. This means, that the excitation spectrum
observed by FIR spectroscopy is not influenced by inter-- dot interaction
and agrees with the one-- electron result (as in  the single dot)}.
This statement is rigorous for parabolic confinement (see Appendix).
It can also be understood intuitively,
 because a $q=0$-- excitation is connected with
homogeneous in-- phase elongations of the dots which do not change
the distance between the electrons. We want to mention that
this conclusion seems to be in contradiction with the experimental
work in Ref.\onlinecite{Kotthaus}. They found a splitting of 
the upper and lower excitation branch at $B=0$ and $q=0$ 
for circular dots in a rectangular
lattice, which they interpreted within a
phenomenological model of interacting dipoles as 
a consequence of lattice interaction. However, they use mesoscopic dots
with a diameter of $370 000 \AA$ and lattice periods of
$400 000$ and $800 000 \AA$. These dots are clearly
beyond our microscopic quantum mechanical model, which rests on a  
parabolic confinement.\\

For the {\em rectangular lattices} considered in our numerical examples
we define
${\bf R}^{(0)}=N_1 a_1 {\bf e}_1+N_2 a_2 {\bf e}_2$ and
${\bf q}=q_1\frac{2 \pi}{a_1} {\bf e}_1+q_2\frac{2 \pi}{a_2} {\bf e}_2$
with the lattice constants $a_1$ and $a_2$ and integers $N_1$ and $N_2$
characterizing the lattice sites.
The components of ${\bf q}$ vary in the Brillouin zone (BZ) in the range
$[-1/2,+1/2]$.
The dipole tensor (\ref{dipole-tensor}) reads
\begin{equation}
{\bf T}(N_1,N_2)=\frac{1}{(N_1^2 a_1^2+N_2^2 a_2^2)^{5/2}}
\left[
\begin{array} {cc}(2N_1^2 a_1^2-N_2^2 a_2^2)&
3N_1N_2a_1a_2\\ 3N_1N_2a_1a_2&(2N_2^2 a_2^2-N_1^2 a_1^2) \end{array}
\right]
\end{equation}
Although for all figures the exact dynamical matrix is used, it
is useful to consider the results with {\em nearest neighbor} (n.n.)
lattice sums in (\ref{dyn-mat}) separately.
This provides simple formulas for order-- of-- 
magnitude estimates. 
\begin{eqnarray}
C_{11}&=&\omega_1^2+2\;p_1\;[1-cos(2\pi q_1)]-p_2\;[1-cos(2\pi q_2)]
\nonumber\\
C_{22}&=&\omega_2^2+2\;p_2\;[1-cos(2\pi q_2)]-p_1\;[1-cos(2\pi q_1)]
\label{C-nn}\\
C_{12}&=&\Omega_{12}
\nonumber
\end{eqnarray}
where we introduced the interaction parameters 
$p_i=\frac{2\beta N}{a_i^3}$.\\[1cm]

\begin{figure}[!th]
\vspace{-2cm}
\begin{center}
{\psfig{figure=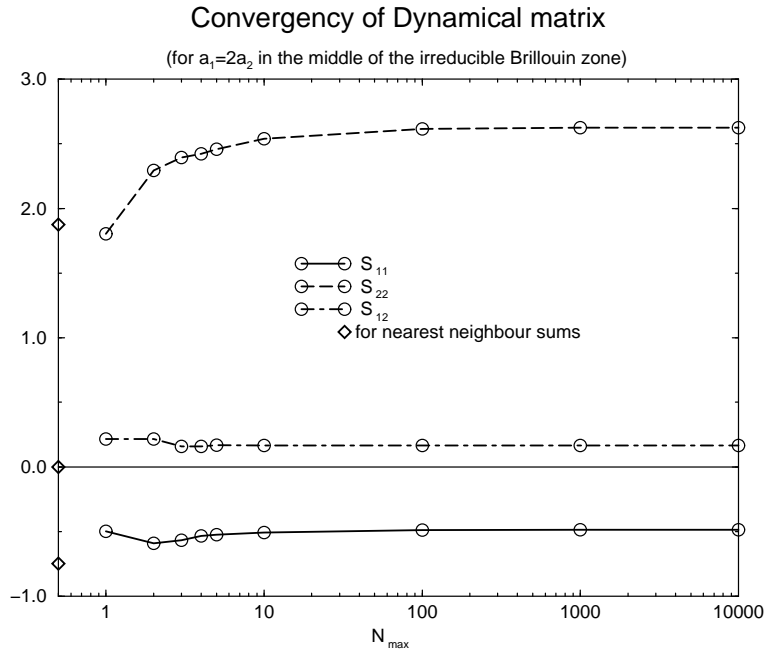,angle=-90,width=12.cm,bbllx=15pt,bblly=45pt,bburx=580pt,bbury=750pt}}
\caption[ ]{
Convergency of the lattice sums $S_{ik}$ as defined in
(\ref{S_ik}) with increasing
number of cubic shells $N_{max}$ for $\bf q$ in the middle of the
Brillouin zone.
}
\label{Fig3}
\end{center}
\end{figure}

The convergence of the lattice sums $S_{ik}$
in the dynamical matrix is shown Fig.3.
$S_{ik}$ is defined by
\begin{equation}
C_{ik}=\Omega_{ik}+p_2\; S_{ik} 
\label{S_ik}
\end{equation}
and depends only on $\bf q$ and the ratio
$a_1/a_2$. Apart from the off-diagonal elements, which vanish in n.n.
approximation, the error of the n.n. approximation is less than $30\%$.

\subsection{Special Features of the Magneto-- Phonon Spectrum}
Fig.s 4-6 show the magneto-- phonon 
\footnote{ The term {\em magneto-- phonon} is attributed to the fact that
the there is no exchange and there are harmonic forces 
between the oscillating individuals. One could also call them
{\em magneto-- plasmons}, if one wants to emphasize that it is only
electrons which oscillate, and no nuclei.}
spectrum for circular dots
on a rectangular lattice with $a_1=2 a_2$.
Because the two interaction parameters have a fixed ratio, it
suffices to use one of them for characterizing the interaction strength.
We chose the larger one $p_2=p$.
For $B=0$ and isolated dots $(p=0)$, the two excitation modes 
are degenerate. If we tune up the interaction strength represented by $p$,
a $\bf q$ dependent splitting appears (see Fig. 4). \\[1cm]

\begin{figure}[!th]
\vspace{-2cm}
\begin{center}
{\psfig{figure=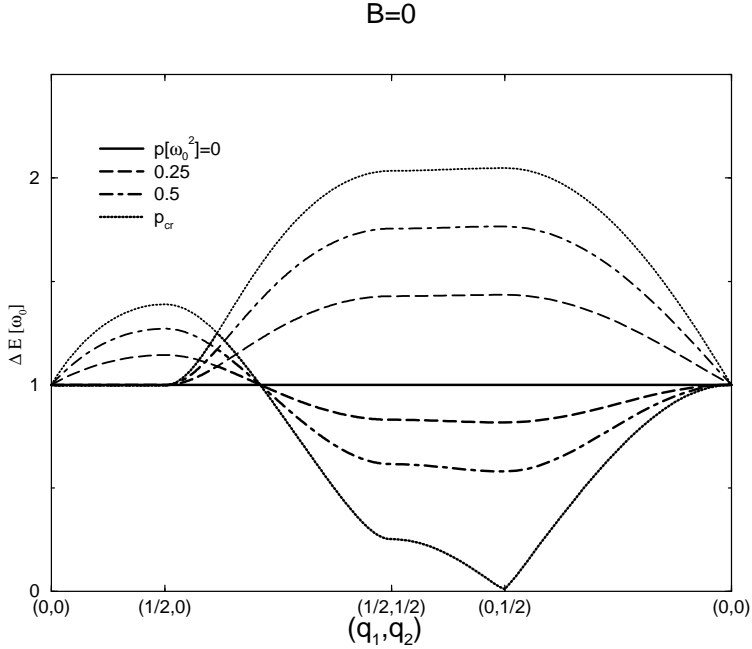,angle=-90,width=12.cm,bbllx=15pt,bblly=45pt,bburx=580pt,bbury=750pt}}
\caption[ ]{
Magneto-- phonon dispersion at $B=0$ for several interaction parameter
values and $\bf q$ on symmetry lines of the Brillouin zone.
($p=p_2$ values in the inset are in units $\omega_0^2$.)
Thick (lower) and thin (upper) lines indicate $\Delta E_-$ and $\Delta E_+$,
respectively.
}
\label{Fig4}
\end{center}
\end{figure}

This splitting is a manifestation of the dot interaction.
For a certain critical
$p_{cr}$ the lower mode becomes soft. This feature will be discussed below.
There are points in the BZ, however, where the {\em degeneracy}
for finite $p$ remains. These
points will be investigated now. We demonstrated in Sect.II after formula
(\ref{B=0}) that {\em necessary} 
for degeneracy is $C_{12}=0$, i.e., the dynamical 
matrix must be diagonal. Then the points with degeneracy are defined
by the condition $C_{11}=C_{22}$.
As seen in (\ref{dyn-mat}),
for circular dots $\omega_1=\omega_2=\omega_0$ this happens
in the center of the BZ ${\bf q}=0$. 
The next question to be discussed is if there are  other points with
degeneracy. The first condition $C_{12}=0$, 
is fulfilled for all points on the surface of the BZ. The second condition
must be investigated for special cases. We find, that for quadratic
lattices $a_1=a_2$ with circular dots $\omega_1=\omega_2$ both modes
are degenerate at the point ${\bf q}=(1/2,1/2)$. In the case shown in Fig.4
this point is somewhere between $(1/2,1/2)$ and $(1/2,0)$. \\
In n.n. approximation (\ref{C-nn}), however, this equation is even 
fulfilled on  full curves in the BZ
defined by \mbox{$p_1\;[1-cos(2\pi q_1)]=p_2\;[1-cos(2\pi q_2)]$}.
In a cubic lattice, this is the straight lines $q_2=\pm q_1$.
The contributions beyond n.n.s remove the exact degeneracy on this curve
in the interior of the BZ,
but leave a kind of anti-crossing behavior of the two branches.\\

An important parameter, which characterizes 
the influence of the dot interaction in circular dots,
is the {\em band width} at $B=0$, i.e. the maximum
splitting of the two branches due to  dot interaction. 
(Remember that this splitting vanishes for noninteracting 
circular dots.)
Assume $a_1> a_2$. Then the largest splitting for circular dots
in n.n. approximation
 appears at ${\bf q}=(0,1/2)$
and has the amount 
\begin{equation}
W=max(\Delta E_+ - \Delta E_-)=max(\omega_+-\omega_-)=\sqrt{\omega_0^2+4\;p_2}-
                  \sqrt{\omega_0^2-2\;p_2}
\end{equation}
For small dot interaction and in units $\omega_0$, this is 
proportional to the interaction parameter
$ \frac{W}{\omega_0} \rightarrow 3 p_2 $.\\

We next discuss the appearance of {\em soft modes}. 
The question is, for which $\bf q$, $B$ and interaction parameter $p$
this happens.
The general condition
for vanishing of the lowest mode is 
$C_{11}\cdot C_{22}=C_{12}^2$ (see Sect. II). 
In this condition the magnetic field does not appear.
For circular dots and with the definition 
(\ref{S_ik}) this equation reads
\begin{equation}
[\omega_0^2+p_2\;S_{11}][\omega_0^2+p_2\;S_{11}]=p_2^2\;S_{12}^2
\end{equation}
After introducing a dimensionless critical interaction parameter
$P_{cr}=p_2/\omega_0^2$, we obtain a quadratic equation for $P_{cr}$
which has the solution
\begin{equation}
P_{cr}=-\frac{1}{2}\frac{tr}{det}
\pm \sqrt{\left( \frac{1}{2}\frac{tr}{det}\right)^2-\frac{1}{det}}
\end{equation}
where $det=S_{11}S_{22}-S_{12}^2$ and $tr=S_{11}+S_{22}$. For 
our numerical  case $a_1=2\;a_2$ and n.n. interaction for $S_{ik}$
the lowest mode becomes soft at ${\bf q}=(0,1/2)$ and the critical interaction
parameter is $P_{cr}=1/2$. Inclusion of lattice contributions beyond n.n.
shifts $P_{cr}$ to 0.7543. The most  important result of this paragraph is 
that lattice softening is {\em independent}
of $B$ (see also Fig.s 5 and 6). 
The latter conclusion is {\em exact} within the range of validity of
 the Hamiltonian (\ref{H-cm-latt}) and no consequence of any
subsequent approximation or specialization.\\[1cm]

\begin{figure}[!th]
\vspace{-2cm}
\begin{center}
{\psfig{figure=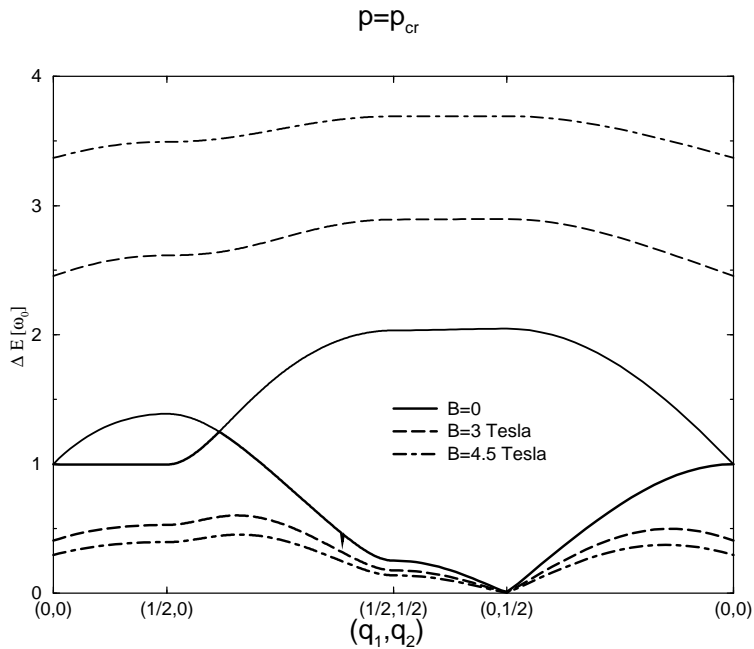,angle=-90,width=12.cm,bbllx=15pt,bblly=45pt,bburx=580pt,bbury=750pt}}
\caption[ ]{
The same as Fig.4, but for $p=p_{cr}$ and several magnetic fields.
}
\label{Fig5}
\end{center}
\end{figure}

\begin{figure}[!th]
\vspace{-2cm}
\begin{center}
{\psfig{figure=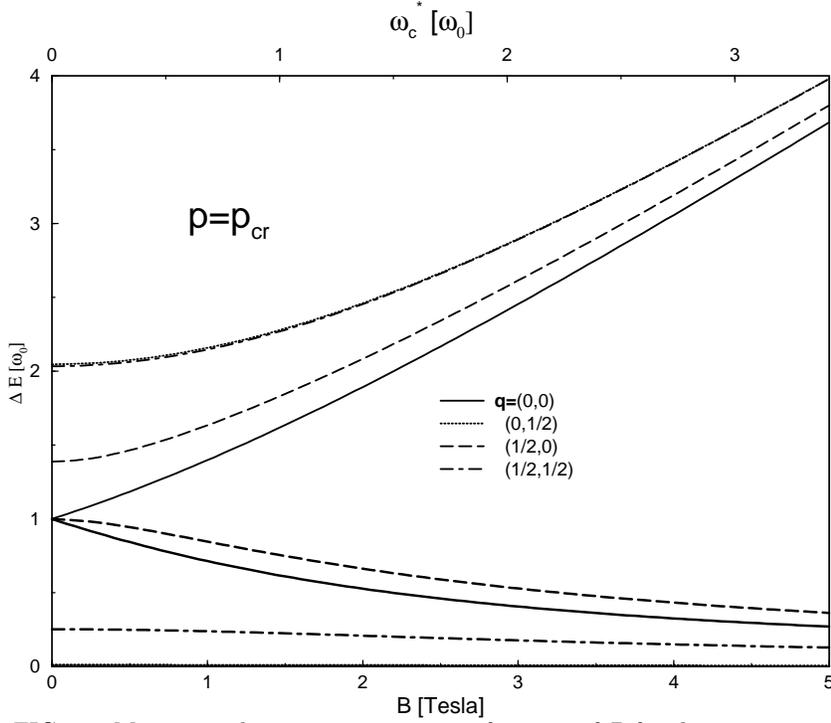,angle=-90,width=12.cm,bbllx=15pt,bblly=45pt,bburx=580pt,bbury=750pt}}
\caption[ ]{
Magneto-- phonon excitations as a function of $B$ for the
symmetry points in the Brillouin zone. The upper abscissa is independent
of the effective mass, the lower one applies to GaAs.
}
\label{Fig6}
\end{center}
\end{figure}

\section{Intra-dot-- Excitations for N=2}
Intra-dot excitations  for circular dots
in a cubic lattice and for $N=2$ can be calculated easily.
We define the relative coordinate ${\bf r}={\bf r}_2-{\bf r}_1$,
and assume that all dots are equivalent (also with respect to
their environment). Then the indexes $(n,\alpha)$ can be chosen as $(0,0)$
and omitted.
The relative Hamiltonian reads 
\begin{equation}
H_{rel}=2 \bigg\{ 
\frac{1}{2 m^*}
\left[
{\bf p}+
{1\over 2 c} {\bf A}({\bf r})
\right]^2 +
\frac{1}{2}\; {\bf r} \cdot {\bf D } \cdot {\bf r}
+\frac{\beta}{2\;r}
\bigg\}
\label{H-rel}
\end{equation}
where
${\bf p}=-i \nabla_{\bf r}$ and
\begin{equation}
{\bf D}=\frac{1}{4}{\bf \Omega}+\frac{\beta}{2}
{\bf T}_0~;~~~~
{\bf T}_0=\sum_{n,\alpha(\ne 0,0)}
{\bf T}\left({\bf R}_{n,\alpha}^{(0)}\right)
\end{equation}
It is worth emphasizing that $H_{rel}$ contains a contribution from
neighboring dots, originating from the interdot Coulomb interaction.
A trivial angular dependent part  can only be decoupled 
from $H_{rel}$, or, the 2-dimensional Schr\"odinger
equation can be traced back to an ordinary radial Schr\"odinger equation,
  if the term ${\bf r} \cdot {\bf D } \cdot {\bf r}$
has the same circular symmetry as the intra-dot Coulomb term
$\beta/(2 r)$. Therefore we confine ourselves to circular dots 
on a cubic lattice,
and we have
\begin{equation}
{\bf T}_0=\frac{1}{a^3}\sum_{N_1,N_2\ne 0,0}\;
\frac{1}{(N_1^2 +N_2^2 )^{3/2}}\;{\bf I}\approx \frac{4}{a^3}\;{\bf I}
\end{equation}
where the simple result is in n.n. approximation.
Using the interaction parameter $p=2N\beta/a^3$ (with $N=2$) defined above,
we obtain
\begin{equation}
{\bf D}=\frac{1}{4}(\omega_0^2+2p)\;{\bf I}
\end{equation}
In this way, dot interaction defines
an effective  confinement frequency 
$\omega_{0,eff}^2 = \omega_0^2+2 p$. This means
that {\em the c.m. excitations have to be calculated (or interpreted) 
with another confinement potential then the relative excitations}.
In our figures we present results for  $\omega_{0,eff}=0.2\; a.u.^*$,
which agrees with the bare confinement potential used in Sect.IV and
the mean value in Sect.III.
Because our results are presented in units of $\omega_0$, they
depend on $\omega_0$ only weakly 
through the differing influence of electron-electron
interaction. For the absolute values, however, the influence
of the dot interaction can be tremendous.\\
In the relative motion there is a coupling between orbital and spin parts
through the Pauli principle. For $N=2$ and a circular effective
confinement, Pauli principle demands that orbital states with 
even and odd relative angular momentum $m_i$
must be combined with singlet and triplet spin states, respectively
( see e.g. Ref.\onlinecite{Taut2e}).
For the c.m. motion there is no interrelation 
between orbital and spin part because the c.m. coordinate
is fully symmetric with respect to particle exchange. 
Consequently, any c.m. wave function can be combined with
a given spin eigen function.
The only spin dependent term in the total energy 
considered here is the {\em Zeeman term}, which reads in our units
\begin{equation}
\frac{E_B}{\omega_0}= 0.9134\cdot10^{-2}\;g_s\;\frac{B[Tesla]}{\omega_0[a.u.^*]}
\;\frac{M_s}{2}
\end{equation}
where we used $g_s=-0.44$ for the gyro-magnetic factor of GaAs 
from Ref.\onlinecite{Merkt}. The 
total spin quantum number is $M_s=0$ for the
singlet state and $0,\pm1$ for the triplet state.
One of the most interesting points in quantum dot physics is that
the total orbital angular momentum of the ground state depends on 
the magnetic field
(see e.g. Ref.s\onlinecite{Merkt},\onlinecite{Maksym}). 
This feature is a consequence of electron-- electron interaction. 
For our parameter values,
the relative orbital angular momentum 
of the ground state $m_i$ changes from  
0 to --1, from --1 to --2, and from --2 to --3 
at $B=1.250$, $4.018$, and  $5.005\; Tesla$.
This corresponds to a sequence $M_s$=0,+1,0,+1 for the spin 
quantum number.
Figures 7a-c show the excitation frequencies for three B-values
lying within the first three regions. 
$m_f$ is the relative orbital
angular momentum of the final state. 
All excitations are
included irrespective of selection rules. 
For dipole transitions only two of them would remain (the lowest excitation
 with $m_f=m_i\pm 1$). For $B=0$, the lowest excitation energy
(in units $\omega_0$) for noninteracting electrons would be 1.
As seen in in Fig.7a, electron-- electron interaction decreases this
value by at least a factor of $1/2$. The same holds qualitatively 
for finite $B$. This is connected to the fact, that the ground 
state depends on $B$. Let's consider an example. For $B=1.250 \;Tesla$
the ground state switches from $m_i=0$ to $m_i=-1$. This implies that
for $B$ approaching this transition field from below, the
excitation energy for
dipole allowed transition from $m_i=0$ to $m_f=-1$ converges to $0$.
In other words, there is a level crossing at the the transition field.
Therefore, very small transition energies and switching of the ground state
are connected.\\[1cm]

\begin{figure}[!th]
\vspace{-2cm}
\begin{center}
{\psfig{figure=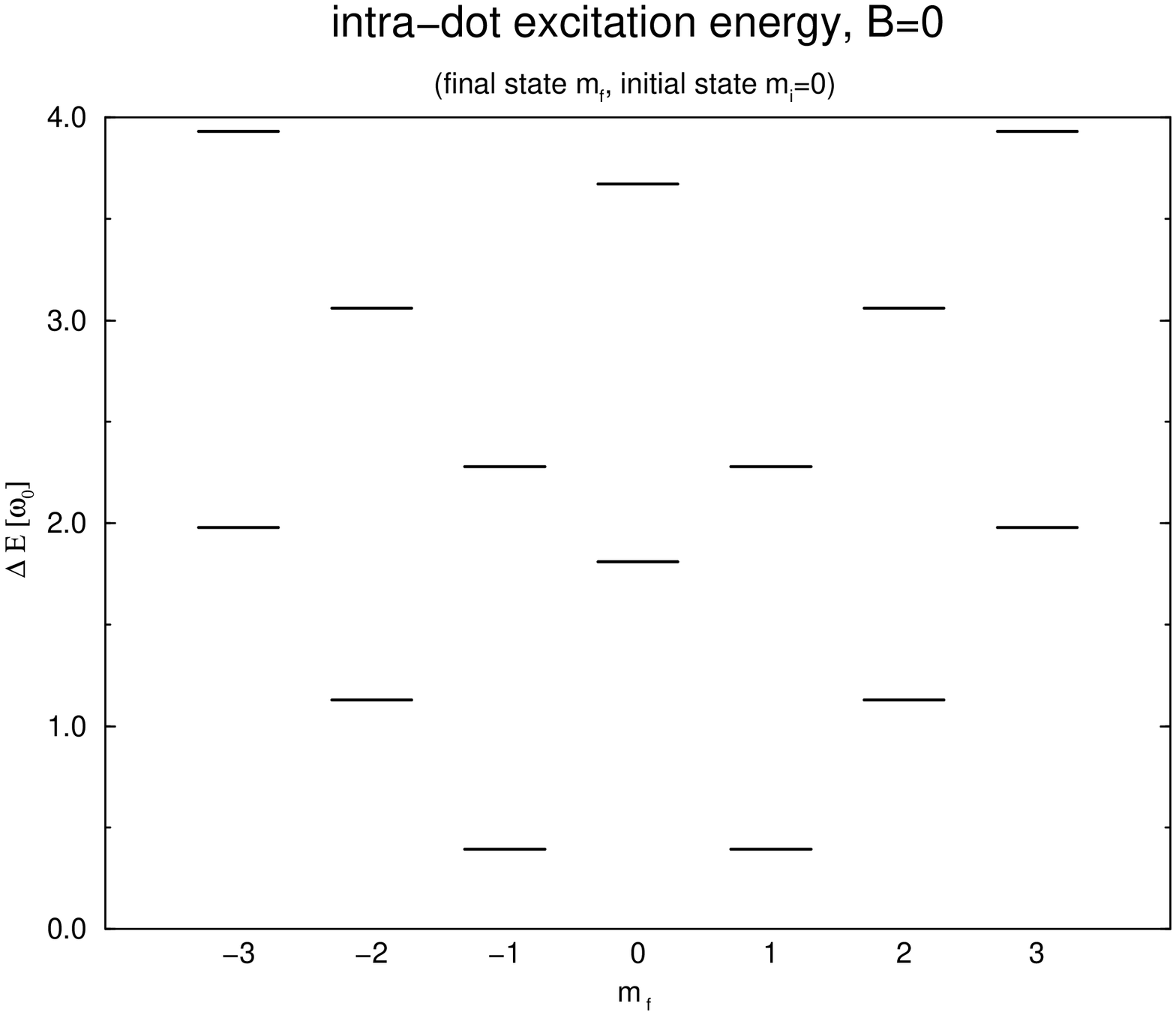,angle=-90,width=12.cm,bbllx=15pt,bblly=45pt,bburx=580pt,bbury=750pt}}
{\psfig{figure=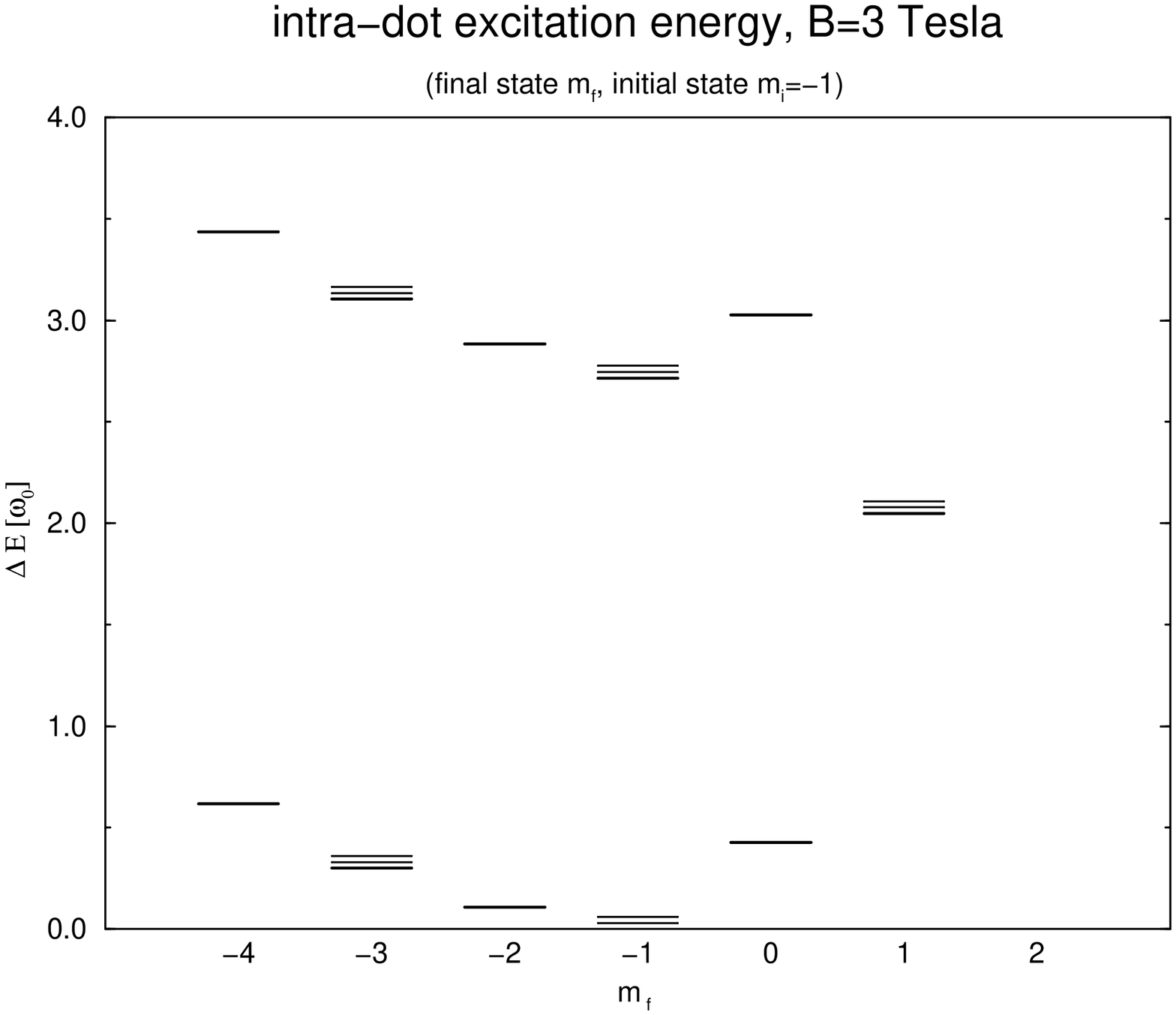,angle=-90,width=12.cm,bbllx=15pt,bblly=45pt,bburx=580pt,bbury=750pt}}
{\psfig{figure=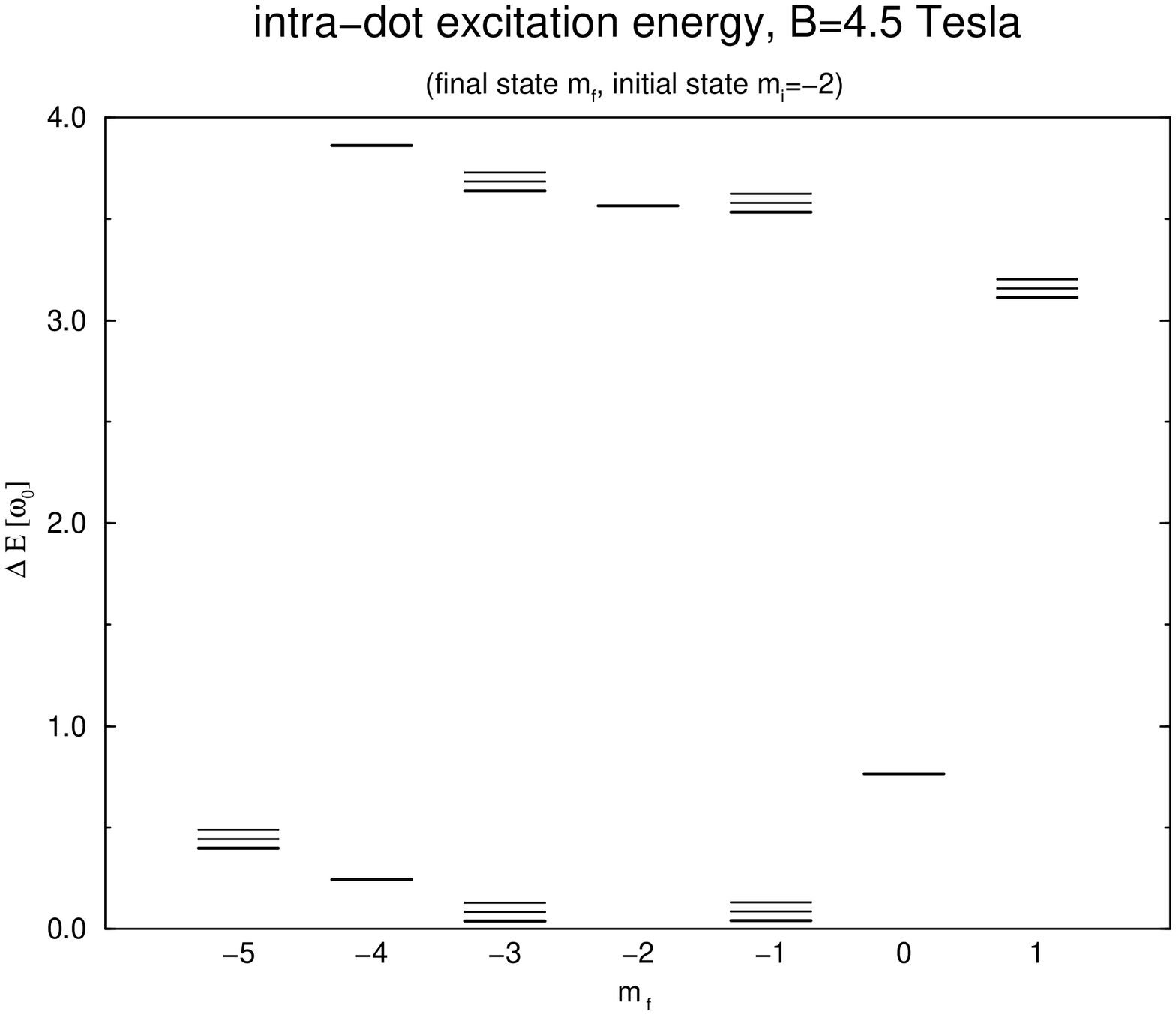,angle=-90,width=12.cm,bbllx=15pt,bblly=45pt,bburx=580pt,bbury=750pt}}
\caption[ ]{
Intradot excitation energies (in units of the {\em effective}
confinement frequency) for B=0 (a), B=3 Tesla (b) and
B=4.5 Tesla (c). The corresponding relative orbital angular momenta
of the ground state are $m_i$=0 (a), --1(b), and --2(c) and
the spin angular momenta $M_s$=0 (a), +1(b), and 0(c).
}
\label{Fig7c}
\end{center}
\end{figure}

For a qualitative understanding, Figures 7a-c
can be used together with Figures 1a, 1b, 4, and 5 to investigate
the relative position of collective  and intra-dot excitations. 
The conclusion is that for small dot interaction (for $p$ well below
$p_{cr}$), the lowest intra-dot excitation energies
lie well below the lowest c.m. excitations. 
Apart from using a different terminology,
this conclusion agrees with the experimental findings in 
Ref.\onlinecite{Heitmann}.\\
Fig.s 7b and c, which belong to finite $B$, show the Zeeman splitting.
All transition energies to final states with odd $m_f$ are triplets because
the corresponding spin state is a triplet state. The thin lines of
a triplet belong to spin-- flip transitions.\\ [.5cm]

\begin{figure}[!th]
\vspace{-2cm}
\begin{center}
{\psfig{figure=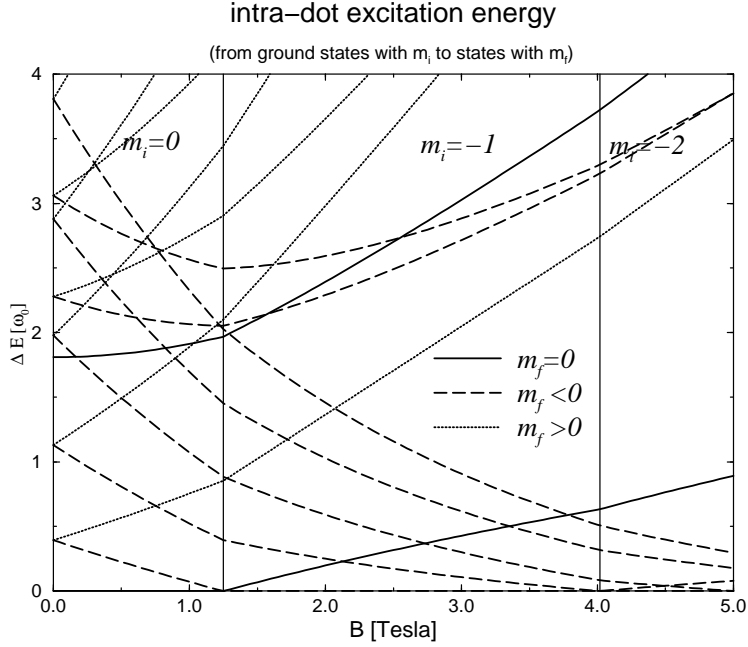,angle=-90,width=12.cm,bbllx=15pt,bblly=45pt,bburx=580pt,bbury=750pt}}
\caption[ ]{
Intradot excitation energies (in units of the {\em effective}
confinement frequency) as a function of the magnetic field.
The small Zeeman splitting is neglected.
The B-- values, where the angular momenta of the ground state
change, are indicated by vertical lines. The absolute value of the
final state  orbital momentum $m_f$  of the curves at B=0
grows from bottom to top by 1 starting with 0.
}
\label{Fig8}
\end{center}
\end{figure}

In Fig.8 the $B$-- dependence of the lowest  excitation energies is shown.
It is clearly seen that the curves exhibit a kink at those $B$-- values,
where the ground state configuration changes. The size of the kink 
decreases with increasing $B$. If this kink could be resolved experimentally
(e.g. by electronic Raman spectroscopy), it would be a direct
indication for the change of the ground state configuration, 
and thus an experimentally observable 
consequence of electron-- electron interaction.

\newpage

\section{Summary}
We solved the Schr\"odinger equation for a lattice 
of {\em identical} parabolic (but not necessarily circular) 
quantum dots with Coulomb
interaction (in dipole approximation) between the dots. 
We provide an overview over the state of art
of these systems which includes the results of former publications.
References can be found in the text.
  
\begin{itemize}
\item
Similar to single dots, the center of mass coordinates of all 
dots can be separated from the relative coordinates.
Only the c.m. coordinates of different dots are coupled to each other. 
The relative coordinates
of different dots are neither coupled to each other nor to the c.m. coordinates.
\item
This gives rise to two types of excitations: two collective c.m. modes
per dot 
and and a complex spectrum of intra-dot excitations.
In periodic arrays only the collective c.m. modes show dispersion.
Intra-dot excitations are dispersion-less.
\item The c.m. system can be solved exactly and analytically
providing magneto-- phonon excitations characterized by a certain
wave number $\bf q$ within the Brillouin zone.
For ${\bf q}=0$ and one dot per unit cell, interdot interaction does not 
have any influence on the c.m. excitations.
\item All dipole allowed excitations (seen in FIR experiments) are not
influenced by the dot interaction.
\item Interdot interaction between two dots influences the spectrum through
a single parameter $p=2 N \beta/a^3$, where $a$ is the  distance between
the dots, $N$ the number of electrons per dot and $\beta$ the inverse
background dielectric constant.
\item If $p$ exceeds a certain critical value $p_{cr}$, the lowest c.m. mode 
becomes soft leading to an instability. This transition is independent of
the magnetic field.
\item For B=0 and and one circular dot per unit cell, the two
c.m. modes are not only degenerate in the middle of the Brillouin zone, 
but also at some points on the surface. If we use the n.n. approximation
 for the
lattice sums in the dynamical matrix, degeneracy is maintained even on
full curves in the Brillouin zone.
\item Intra-dot excitations have to be calculated from an effective
confinement. 
In circular dots with a cubic environment in
nearest neighbor approximation the effective confinement frequency reads
 $\omega_{0,eff}^2 = \omega_0^2+2 p$. 
This effective confinement differs from that  for the
c.m. motion.
\item For $p$ well below $p_{cr}$, the lowest intra-dot excitations
are much smaller than the lowest collective excitations.
\item The intra-dot excitation energies versus magnetic field
exhibit kinks at those fields, where the angular momentum of the 
ground state changes.
\end{itemize}
In the Appendix we prove a  Kohn Theorem for dot arrays with 
Coulomb interaction between the dots without the dipole approximation. 
The individual confinement potentials can be arbitrarily arranged and can carry 
different electron numbers, but have to be described by identical
confinement tensors. This means that for breaking Kohn's Theorem in
dot arrays, we have to have at least two different confinement species.

\section*{Appendix}
We are going to prove that for an arbitrary
array \footnote{The dot centers can be arranged arbitrarily.}
 of identical parabolic quantum dot potentials
\footnote{The confinement tensors $\bf \Omega$ 
of all dots must be equal.}
 in an homogeneous
magnetic field: i) the total c.m. degree of freedom
can be separated from the rest, ii) the total c.m. Hamiltonian 
is not influenced 
by Coulomb interaction,
and iii) the eigenvalues of the total c.m. Hamiltonian 
are independent of the electron number $N$ in each dot \footnote{
The electron number in different dots can be different.} . 
\\
The Hamiltonian $H=H^{(0)}+V$ consists of an one--particle term $H=H^{(0)}$
and the Coulomb interaction between all electrons $V$. The dot centers are
located at ${\bf R}_{\alpha}^{(0)}$ and the electron coordinates are denoted by
${\bf r}_{i \alpha}={\bf R}_{\alpha}^{(0)}+{\bf u}_{i \alpha}$. Then we have
\begin{equation}
H^{(0)}=\sum_{i \alpha}
\left\{
\frac{1}{2 m^*}
\left[
{\bf p}_{i \alpha}+
{1\over c} {\bf A}({\bf R}_{\alpha}^{(0)}+{\bf u}_{i \alpha})
\right]^2 +
\frac{1}{2} \;{\bf u}_{i \alpha} \cdot
{\bf C }
\cdot {\bf u}_{i \alpha}
\right\}
\label{H-0}
\end{equation}
First of all we shift the gauge center for each electron into the middle
of the corresponding dot using an unitary transformation similar to
(\ref{gauge-shift}). 
This transforms the shift ${\bf R}_{\alpha}^{(0)}$ in the argument 
of the vector potential away.
Next we drop the index $\alpha$ in (\ref{H-0})
 so that the index
'$i$' runs over all electrons in all dots. Now we perform  a
transformation to new coordinates $ \tilde {\bf u}_i$
\begin{equation}
{\bf u}_i=\sum_k Q_{ik}\;(\sqrt{N}\; \tilde {\bf u}_k)~;~~~
(\sqrt{N} \;\tilde {\bf u}_i)=\sum_k Q_{ki}^* \; {\bf u}_k
\end{equation}
where $Q_{ik}$ is an unitary matrix. This implies
\begin{equation}
{\bf p}_i=\sum_k Q_{ik}^*\;(\frac{\tilde {\bf p}_k}{\sqrt{N}} )~;~~~
(\frac{\tilde {\bf p}_i}{\sqrt{N}} )=\sum_k Q_{ki} \; {\bf p}_k
\end{equation}
It is possible to choose for the first column 
$Q_{k1}=\frac{1}{\sqrt{N}}$. The other
columns need not be specified. 
Then $\tilde {\bf u}_1=(1/N)\sum_i{\bf u}_i={\bf U}$ is the c.m.
of all elongations, or, the c.m. of the electron coordinates
with respect to the weighted center of the dot locations
${\bf R}^{(0)}=(1/N)\sum_\alpha N_\alpha\;{\bf R}_{\alpha}^{(0)}$,
where $N_\alpha$ is the number of electrons in dot $\alpha$.
The corresponding canonical momentum 
$\tilde{\bf p}_{1}=(1/i)\nabla_{\tilde {\bf u}_1}={\bf P}$
is the c.m. momentum.
Inserting our transformation into (\ref{H-0}) provides
\begin{equation}
H^{(0)}=\sum_i
\left\{
\frac{1}{2 m^*}
\left[
\frac{1}{\sqrt{N}} \;\tilde{\bf p}_{i}+
{\sqrt{N}\over c} {\bf A}(\tilde{\bf u}_{i})
\right]^2 +
\frac{N}{2} \;\tilde{\bf u}_{i} \cdot
{\bf C }
\cdot \tilde{\bf u}_{i}
\right\}
\label{H-0-trans}
\end{equation}
The term $i=1$ in (\ref{H-0-trans}) is the (separated) c.m. 
Hamiltonian 
\begin{equation}
H_{c.m.}=\frac{1}{2 m^*}
\left[
\frac{1}{\sqrt{N}}\; \tilde{\bf P}+
{\sqrt{N}\over c} {\bf A}(\tilde{\bf U})
\right]^2 +
\frac{N}{2} \;\tilde{\bf U} \cdot
{\bf C }
\cdot \tilde{\bf U}
\end{equation}
which agrees with (\ref{H-cm}). Clearly, the Coulomb interaction
$V$ in $H$ is independent of the c.m., and does not contribute to
$H_{c.m.}$.
For the independence of the eigenvalues of
$N$ see the discussion following (\ref{H-cm}).\\

This prove,  in particular
the step from (\ref{H-0}) to (\ref{H-0-trans}),
is not correct if the dot confinement tensor $\bf C$ depends
on $\alpha$ (or '$i$' in the changed notation). Therefore, all dots
must have the same $\bf C$, but can have different electron numbers
$N_\alpha$. In other words, the total c.m. excitations in dot arrays, 
which are seen in
FIR spectra, are not affected by the e-- e-- interaction, if and only if
all confinement tensors $\bf C$ are equal. On the other hand, {\em if we want
to observe e-- e-- interaction in the FIR spectra and  break
Kohn's Theorem, we have to use  dot lattices  with at least two different
confinement tensors}. The simplest way to implement this is using
a lattice with two non-circular dots per cell, which are 
equal in shape, but rotated
relative to each other by 90 degrees.

\section{Acknowledgment}

I am indebted to  D.Heitmann, H.Eschrig, and E.Zaremba and their groups
for very helpful discussion and the
Deutsche Forschungs-- Gemeinschaft
for financial funding.

\end{document}